\documentclass[a4paper]{spie}  
\usepackage[]{graphicx}
\usepackage{hyperref}
\usepackage{aas_macros}

\title{The numerical simulation tool for the MAORY multiconjugate adaptive optics system} 

\author{
Arcidiacono C.\supit{a}, Schreiber L.\supit{a}, Bregoli G.\supit{a}, Diolaiti E.\supit{a}, Foppiani I.\supit{a}, Agapito G. \supit{d} , Puglisi A. \supit{d}, Xompero M.\supit{d}, Oberti S.\supit{e}Cosentino G.\supit{b}, Lombini M.\supit{a}, Butler R. C.\supit{c} Ciliegi P.\supit{a}, Cortecchia F. \supit{a}, Patti M. \supit{b}, Esposito S.\supit{d} and Feautrier P.\supit{f}
\skiplinehalf
\supit{a}INAF - Osservatorio Astronomico di Bologna, Via Ranzani 1, I-40127 Bologna, Italy; \\
\supit{b}Universit\'a degli Studi di Bologna, Dipartimento di Fisica e Astronomia, Viale Berti Pichat, 6/2 I-40127 Bologna, Italy;\\
\supit{c}INAF - Istituto di Astrofisica Spaziale e di Fisica Cosmica, via Piero Gobetti 101, I-40129 Bologna, Italy\\
\supit{d}INAF - Osservatorio Astrofisico di Arcetri, Largo Enrico Fermi 5, I-50125 Firenze, Italy;\\
\supit{e}ESO  - European Southern Observatory, Karl Schwartzchild strasse 2, D-85748 Garching bei M\"unchen, Italy;\\
\supit{f}Institut de Planétologie et d'Astrophysique, 414 Rue de la Piscine, 38400 Saint-Martin-d'Hères, France
}

\authorinfo{Further author information: (Send correspondence to Carmelo Arcidiacono)\\C.A.: E-mail: carmelo.arcidiacono@oabo.inaf.it, Telephone:  +39 051 2095 704}

\begin{document} 
  \maketitle 

\begin{abstract}
The Multiconjugate Adaptive Optics RelaY (MAORY)  is and Adaptive Optics module to be mounted on the ESO European-Extremely Large Telescope (E-ELT). It is a hybrid Natural and Laser Guide System that will perform the correction of the atmospheric turbulence volume above the telescope feeding the Multi-AO Imaging Camera for Deep Observations Near Infrared spectro-imager (MICADO).
We developed an end-to-end Monte-Carlo adaptive optics simulation tool to investigate the performance of a the MAORY and the calibration, acquisition, operation strategies.
MAORY will implement Multiconjugate Adaptive Optics combining Laser Guide Stars  (LGS) and Natural Guide Stars (NGS) measurements. 
The simulation tool implements the various aspect of the MAORY in an end to end fashion. The code has been developed using IDL and use libraries in C++ and CUDA for efficiency improvements. Here we recall the code architecture, we describe the modelled instrument components and the control strategies implemented in the code.  
\end{abstract}



\section{INTRODUCTION}
\label{sec:intro}  
The optical turbulence above ground-based telescopes limits the achievable spatial resolution. 
The Adaptive optics (AO) is a technology developed to enable high-resolution imaging from the ground. 
The concept of adaptive optics compensation is more than 50 years old\cite{1953PASP...65..229B,Babcock253} and it is working on sky since 1989\cite{1990A&A...230L..29R}. 
It simply foresees to compensate in real-time for the phase perturbations introduced by the atmospheric turbulence.
The simplest configuration of an AO systems is
the single-conjugate. A single-conjugate adaptive optics (SCAO) system presents
a (one) deformable mirror (DM) to compensate , and a (one) single wavefront sensor (WFS)
to measure the phase perturbation residuals in the direction of the reference source.
On the ESO - Multiconjugate Adaptive Optics Demonstrator (MAD)\cite{hubin02,2003SPIE.4839..317M,marchetti05} the multi deformable mirror version, called Multi-Conjugate Adaptive Optics\cite{beckers88,beckers89a} (MCAO), has been demonstrated on sky. In MCAO using many reference sources, it's possible to obtain a correction of the optical turbulence above the telescope valid for a Field of View and not only for the special direction of the reference star as in SCAO case.
Using two deformable mirrors and three Natural Guide Stars (NGS) in the Star Oriented Approach\cite{marchetti06} or three to eight NGS in the 
Layer-Oriented\cite{LO1,LO2,arcidiacono07} fashion,
a uniform correction for a field of view of 2arcmin has been achieved\cite{2007Msngr.129....8M,2008SPIE.7015E..5PA}.

We developed a numerical simulation tool\cite{2014SPIE.9148E..6FA} for the end to end simulations of the Multiconjugate Adaptive Optics RelaY\cite{maory} (MAORY) for the ESO European-Extremely Large Telescope\cite{2007Msngr.127...11G,E-ELT}. MAORY is the adaptive optics module of the E-ELT that will feed the Multi-AO Imaging Camera for Deep Observations Near Infrared spectro-imager (MICADO) through a gravity invariant exit port. MAORY has been foreseen to implement MCAO correction through three high order deformable mirrors driven by the reference signals of six Laser Guide Stars (LGSs) feeding as many Shack-Hartmann Wavefront Sensors\cite{hartmann}. A three Natural Guide Stars (NGSs) system will provide the low order correction.\\ The simulation tool is based on the IDL language performs a Monte-Carlo modelling of the MAORY system performance through an extensive usage of the available GPUs. Here we recall the code architecture and describe the modelled instrument components and the control strategies we implemented. The NGS system limiting magnitude and its interaction with the LGS adaptive loop is fundamental to consolidate the design of the MAORY.
In February 2016 the Phase B of the MAORY project has started\cite{2014SPIE.9148E..0YD}, in the next two years, the simulation tool will be extensively used for the trade-offs solutions and concepts verifications.

\section{Numerical Simulation Tool}
Given the previous experiences in numerical computing of the MAORY-team, it was decided to use the IDL language 
for the development of the architecture of the code. A set of high level routines take into account for the system configurations, the simulation skeleton and the main loop. The low level routines provide the repetitive and mathematical jobs. Some of the low level routines of the STARFINDER\cite{2000Msngr.100...23D} and of the LOST\cite{2004ApOpt..43.4288A} numerical tools have been used and adapted to serve the new tool.\\
The End-to-End simulations of Extremely large telescopes (ELTs) foresee extreme performances to get the output in a reasonable short time. The use of multi--core workstation is mandatory and the requirements on memory are challenging: at least 128Gb memory RAM are needed for a full MCAO ELT simulation.
An End-to-End E-ELT configuration requires in the most important moment, the closed loop phase, to store in the RAM mirror modes, open loop wavefront (WF) arrays of the simulated Adaptive Optics (AO) references and test stars (for Point Spread Function (PSF) and Strehl Ratio (SR) computation), control matrix, slope vectors and a few more service arrays and structures. For the MAORY we typically consider a telescope pupil inscribed on 740$\times$740 square array for a 4.875e-2 meter per pixel, this value sets all the other important ones: considering $\approx 5000$  modes for the ground layer Deformable Mirror (DM), $\approx 1400$ and $\approx 1800$ respectively for the two post-focal DMs with 1m actuator pitch projected on the primary, a $3\times 3$ constellation of test stars, for 2seconds of run, the simulation needs about 100GB of available RAM.

Many routines of IDL are multi-threaded, however, a few ones that are particularly used in the code are not. As an example, we give here the case of the Singular Value Decomposition in some of the form available in the IDL library ($\tt{SVDC, LA\_SVD, IMSL\_SVD}$). Moreover, some of the IDL-routines that use the thread pool, such as the {\tt FFT} (actually Discrete Fourier Transform {\tt DFT}) are not performing as fast as freely available solutions such as the {\tt fftw} the Fastest Fourier Transform in the West~\cite{FFTW05}.

\begin{figure}
   \begin{center}
   \begin{tabular}{c}
   \includegraphics[width=6cm]{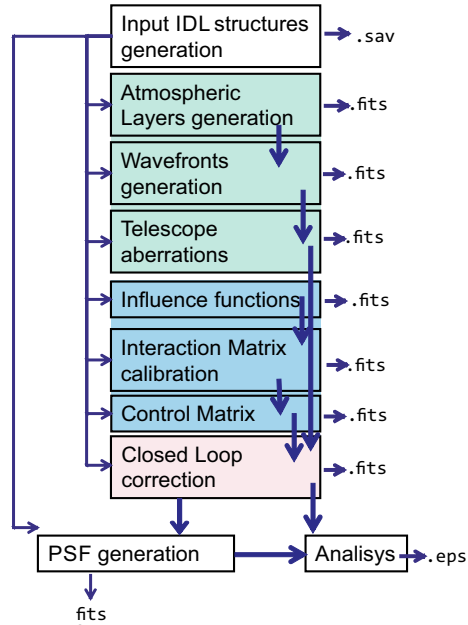}\\
   \end{tabular}
  \end{center}
   \caption[example] 
   { \label{fig:scheme} 
The scheme of the simulation tool.}
   \end{figure}

High-Performance Computing (HPC) is looking more and more to the use of  general purpose graphic processors (GPGPUs). In the recent years, the cost and performance became interesting for a large community of users increasing the availability of mathematical/physical libraries for numerical computation. In particular, we are interested in the use of a GPU to accelerate the numerical computation. \\
NVIDIA set a new standard with TESLA based on the NVIDIA Kepler\texttrademark Architecture for scientific computing, thanks also to the development of the Compute Unified Device Architecture\cite{CUDA} (CUDA). 
NVIDIA created a parallel computing platform for the GPUs they produce, which is almost perfectly compatible with the full C standard. Actually, it runs code through a C++ compiler.\\
From the IDL environment, it's possible to call external library through the dynamically loadable modules ({\tt DLM}). In particular, in this way we have access to parallel optimised C/C++ and CUDA libraries directly from IDL.
We designed the simulation tool following a modular approach, focusing on the possibility to re-utilize (recycle) as much as possible of the already simulated cases: the different simulation steps, namely
the atmosphere generation, the open loop wave-front measurements,
the inclusion of telescope aberrations, the interaction matrix calibration, the closed loop, the Point Spread
Functions (PSFs) generation are performed independently. The results of
each of these steps are saved in fits\cite{fits} files on the disc and these may be the input
for the following modules. See Figure~\ref{fig:scheme}.

A full simulation is then divided in standing alone components which play a specific role in the logic the adaptive loop.
Some of these logical components of the End-to-End simulation may be independent of the closed loop (as the interaction matrix of the open loop) or even from the adaptive optics instrument simulated (for example the atmospheric phase screens).\\
\subsection{The atmospheric layers generation} The optical turbulence is modelled through a set of phase screens representing turbulent layer at different altitudes. Each phase screen is computed by DFT, using the FFTW library, and applying the Kolmogorov\cite{1941DoSSR..30..301K,1941DoSSR..32...16K} or Von Karman\cite{Winker:91}
models for the phase Power Spectrum. The phase screens are completely static following the frozen flow hypothesis. Indeed each phase screen is randomly generated from a set of variable that has a statistical sense such as the coherence length or the outer-scale.
The phase screens arrays are saved on a {\tt .fits}\cite{fits} file, in order to be easily re-used and inspected. 
\subsection{The open loop wave-front measurements}
The code computes from the phase screen the open loop wavefront (WF) for all the stars. Actually, the tool simulates the open loop history as a series of WF for the reference and for the PSF or test star over the field of view of interest. The code computes the WF of the LGS by considering the geometrical projection of the laser guide reference placed at a finite distance from the phase screen placed at the corresponding layer altitude.
The code computes the loop phase history series and it saves it on a {\tt .fits} file for the user defined angular directions and the whole simulated duration of the test.
\subsection{Telescope Aberrations}
On the open loop phase, the code may add a number of telescope aberrations. The telescope aberrations are optical phase screens that are summed to the existing open loop data. The code is able to consider a time series of Zernike polynomial coefficients (to take into account telescope vibration for example). The code uses the possibility to add extra aberrations to mimic the effect of a mispositioning of the Natural Guide Star WFSs (by adding and extra tip and tilt aberration). Other static (or quasi-static) aberration may be piled up to phase: the code reads user-defined input phase map to be added to the open loop WF history of all, or part of, the simulated direction and reference stars. In this way we may consider for example the scallopping error generated by the E-ELT primary.
\subsection{Influence function and modal base}  
The deformable mirrors are computed as the linear combination of a modal base. The actuators base defined by the measured or by the expected influence functions
is the base from which the code starts to compute a modal base. The deformable mirrors are then ready to be used to register the interaction matrix and to close the loops. The tool accepts user-supplied influence functions or it may compute those analytically. They can be used to 
fit Zernike modes or to compute the Karhunen-Loeve\cite{karhunen} expansion which best fits the stochastic wave--fronts induced by atmospheric turbulence. Again, the result may be easily inspected since the computed modes are stored in a (large) {\tt .fits} file.
\subsection{The calibration of the interaction matrices}
The modal base selected as input is used to register the slopes as they were measured by the WFS modules. This procedure may consider possible environment conditions (noise) and simulate the calibration measurement as the were obtained on-sky (using newly generated phase screens). The interaction matrix and the reference slopes vector to be used in closed loop are saved in a {\tt .fits} file.
\subsection{Control Matrix}
The feedback reaction of the system is defined mainly through the computation of the control matrix. It may be easily the (pseudo) inverse of the array built using the DM - WFS interaction matrices and selecting the valid sub-aperture, also known as Least-Square
Estimator (LSE), possibly truncated (TLSE). Or may consider a Bayesian inference approach using the prior knowledge about noise and turbulence statistics to compute a Minimum Mean Square Error (MMSE) estimator. The user may pass a vector of weights to tune the tomographic re-partition of the turbulence on the different DM. Moreover he has the chance also to tune the regularization of the control matrix trough setting a coefficient. 
\subsection{The closed loop}
The core of the simulation is the closed loop phase. The WFS modules compute the vector signals (X- and Y- slopes) from the residual phase (the open loop wavefront subtracted by the DM).
Vector slopes multiplied by the control matrix returns the vector coefficients for the modal base linear composition, in the form of the differential coefficients to be added to the one defining the existing mirror shape. It's possible to consider at this level the sodium layer temporal evolution. 
For the controller, the user may choose to have a pure integrator or a second order low pass filter.
\subsection{The Point Spread Functions (PSFs)}
The user select the wavelength at which the PSF is computed and the pixelscale. This module of the code makes an extensive use of the GPU. The user may desire more wavelengths, considering a bandwidth. Actually the user may decide to compute the Strehl Ratio (SR) on the PSF or on the Close Loop WF history using the Marechal approximation.
\subsection{Analysis}
We developed the first set of tools to allow a quick look at the data and obtain interesting quality merit function from the PSF and from the Closed Loop history file.
Typically we look into the SR map over the FoV and to the residual error time evolution on the various directions considered.
\section{Conclusions}
We are using the simulation tool to perform trade-off analysis and to take confidence with the MAORY system. We plan to study different control strategies and in particular to simulate the possible improvement achievable using the Fractal Iterative
Method\cite{frim} (FrIM) estimator or the pseudo–open loop control\cite{polc} (POLC). Another important update will consider the rotation of the modal bases (the DMs, or some them) with respect to the WFS and the way the MAORY will take care of it. 
\acknowledgments     
The first author wishes to thank all the MAORY consortium for the great work we are doing together. This work has been partly supported by the Italian Ministero dell’Istruzione, dell’Universit\`a e della Ricerca (Progetto Premiale E-ELT 2012 – ref. Monica Tosi).

\bibliography{report}   
\bibliographystyle{spiebib}   

\end{document}